# VoIP Call Optimization in Diverse Network Scenarios Using Learning Based State-Space Search Technique

Tamal Chakraborty[1], Atri Mukhopadhyay[2], Iti Saha Misra[1], Salil Kumar Sanyal[1]

[1]Dept. of Electronics & Telecommunication Engineering, Jadavpur University, Kolkata-700032, India
[2]School of Mobile Computing and Communication, Jadavpur University Salt Lake Campus, Kolkata-700098, India
`{tamalchakraborty29, atri.mukherji11}@gmail.com, iti@etce.jdvu.ac.in,
s_sanyal@ieee.org`

## ABSTRACT

*A VoIP based call has stringent QoS requirements with respect to delay, jitter, loss, MOS and R-Factor. Various QoS mechanisms implemented to satisfy these requirements must be adaptive under diverse network scenarios and applied in proper sequence, otherwise they may conflict with each other. The objective of this paper is to address the problem of adaptive QoS maintenance and sequential execution of available QoS implementation mechanisms with respect to VoIP under varying network conditions. In this paper, we generalize this problem as state-space problem and solve it. Firstly, we map the problem of QoS optimization into state-space domain and apply incremental heuristic search. We implement the proposed algorithm under various network and user scenarios in a VoIP test-bed for QoS enhancement. Then learning strategy is implemented for refinement of knowledge base to improve the performance of call quality over time. Finally, we discuss the advantages and uniqueness of our approach.*

## 1. INTRODUCTION

Voice over Internet Protocol (VoIP) [1] has witnessed rapid growth in recent years owing to ease of network maintenance and savings in operational costs. As it is being widely deployed in office and public networks, maintaining the Quality of Service (QoS) of an ongoing call has assumed utmost importance. Network parameters such as bandwidth, error rate, loss rate, latency, etc. varies with time. With increasing number of users, the issues related to admission control, fairness, scalability, etc also need to be properly addressed. So the QoS optimization techniques must be adaptive.

However, abrupt implementation of these techniques without maintaining proper sequence often results in degraded performance. For example, Random Early Detection (RED) buffer is not advantageous without any end-to-end congestion control mechanism [2]. Further, it is observed that often such abrupt implementations of optimization techniques conflict with each other. For example, RED implementation for small buffer size is not better than static queue with tail-drop mechanism [2]. However, buffer size must be kept small to reduce delay in real-time traffic during congestion. Thus they conflict each other. So the decision to apply appropriate optimization technique is crucial.

This paper aims to generalize the problem of QoS implementation amid diverse scenarios by mapping it as state-space problem. The objective is to maintain adaptive QoS in varying scenarios by application of the most suitable available QoS optimization technique. Focus is also on prioritizing emergency calls with QoS guarantees.

A state space search is the method of finding Goal state(s) from start state through certain intermediate states [3]. In heuristics based search, each state is given a heuristic and traversing is done following a heuristic function. Incremental search further reuses information from previous searches to speed up current search [4]. Incremental heuristic search combines the features of both and is used for developing an algorithm to fulfil the aforementioned objective. The proposed algorithm is implemented in single and multiple voice and video calls in the test-bed to maintain the call quality as network conditions are modified. We extend the work by introducing learning in the proposed algorithm to refine the knowledge base at run-time for improvement in call quality.

The algorithm developed by us is flexible in the sense that any appropriate QoS implementation mechanism can be incorporated after careful analysis. Adaptive strategies such as [5] and [6] and cross layer optimization schemes like [7] are the current areas of research in the field of QoS maintenance and hence can be incorporated in our algorithm after proper examination of the related factors. Moreover the process of continuous learning ensures that any error in the knowledge base during initial development is detected and corrected thereby increasing the efficiency of the algorithm with respect to time.

## 2. MOTIVATION

Formally, a state space can be defined as a tuple [N, A, S, G]
    where, N is a set of states,
        A is a set of arcs connecting the states,
        S is a nonempty subset of N that contains start states, and
        G is a nonempty subset of N that contains the goal states.

There can be various static search strategies like Depth First Search (DFS) [8], Breadth First Search (BFS) [8], etc. While such search strategies can map many real-world problems, at times we need to apply heuristics based search to reflect the dynamics of the problems. Heuristic based search uses domain specific knowledge to choose the successor state and therefore takes into account the varying nature of the problem environment. It is a method that may not always find the best solution but is guaranteed to find a good solution in reasonable time. Incremental search is another search strategy that solves dynamic shortest path problems, where shortest paths have to be found repeatedly as the topology of a graph or its edge costs change [9]. They can find optimal paths to similar problems more easily by reusing information from the previous problems.

Various incremental and heuristic based search methods are being applied to solve problems in the field of symbolic planning [10], [11], path planning in the form of mobile robotics and games [12], [13], reinforcement learning [14], control problems [15], etc. Networking domain has also witnessed the usage of dynamic state-space search techniques for optimal route planning [16]. However, little work has been done with respect to mapping a particular network related problem, other than finding shortest routes, to a state-space problem and solving it. For instance, cell-to-switch assignment is developed using heuristic search for mobile calls in [17]. Heuristic search is further used in [18] to simulate possible attackers searching for attacks in modelled network for proactive and continuous identification of network attacks. Recursive Random Search (RRS) strategy is used in [19] to build an online simulation framework to aid generic large scale network protocol and parameter configuration. Dynamic channel allocation in mobile cellular networks while maintaining QoS has been formulated as state-space problem in [20] and solved by applying heuristic search technique. Similarly, our aim is to address the issue of QoS maintenance in real-time applications like VoIP by mapping it into a state-space search problem.

Real-time constraints on search strategies imply that search must be restricted to a small part of the domain that can be reached from the current state with a small number of action executions [21]. Real-time heuristic search is one such technique that makes planning efficient by limiting the search horizon. Korf's Learning Real-Time A* (LRTA*) method [22] is probably the most popular real-time search method used. Another technique is incremental heuristic search, which makes planning efficient by reusing information from previous planning episodes to speed up the current one as described in D* Lite [23] and Lifelong Planning A* (LPA*) [24] algorithms. Thus real-time search techniques with an upper bound on planning and execution time are considered suitable for implementation in real-time communication domain like VoIP.

As optimizations in the field of QoS maintenance continue to evolve and mitigate the effects of unpredictable nature of networks, implementing them in proper sequence to achieve the highest performance efficiency is the biggest challenge. Moreover, each such QoS implementation mechanism must be maintained adaptively to cope up with variations in the network or changes in the user scenarios. To resolve conflicts, the system must capture system policies, including end-to-end scheduling policies, policies to decide which application's QoS to degrade when there are not enough resources to provide the desired QoS to all applications, and admissions control policies [25]. Our objective is, therefore, to propose an optimization algorithm driven by real time learning based heuristic incremental state–space search that fulfils the aim of maintaining adaptive QoS in multiple call scenarios and under diverse network conditions by applying the available QoS optimization techniques in proper sequence. Thus our work is based on two principal steps namely,
   1. Optimization of VoIP call using Dynamic Search, and
   2. Implementation of Learning Strategy for QoS Enhancement.

## 3. OPTIMIZATION OF VoIP CALL USING DYNAMIC SEARCH

The aim is to map the problem of optimizing VoIP over various network links into a state-space domain where the next state from a set of intermediate states is selected based on incremental heuristic search obeying certain constraints.

### 3.1. Proposed Mechanism

The state-space is defined as a tuple [N, A, S, G].
- 'S' contains the start state which is defined as the call initiation state with respect to time and having heuristics namely delay, loss and Mean Opinion Score (MOS).
- 'G' contains the call termination state as the Goal state with respect to time along with its related heuristics as stated above.
- 'N' contains all the intermediate states within. An intermediate state is taken as any part of an ongoing call with respect to time along with its related heuristics. Heuristics can be categorized as *Excellent*, *Good*, *Average* and *Poor* based on user satisfaction level. Any intermediate state is derived by variation in the network parameters, significant change in heuristic values and by application of QoS optimization techniques.
- 'A' is a set of arcs from one state to another and is effected by transition functions namely $\delta_1$, $\delta_2$ and $\delta_3$. $\delta_1$ is network triggered and can occur due to changes in network. $\delta_2$ is performed by the user in response to $\delta_1$ and involves applying QoS optimization techniques. $\delta_3$ is again user triggered and maintains QoS in a multiple call scenario.

Every heuristic must obey certain local constraints for each call. Delay and loss must be within 180 ms and 5% respectively. MOS must be at least 2. In multiple call scenario, global constraints are taken as mean of local constraints. Now the proposed algorithm is discussed. It consists of two phases, namely analysis and implementation. Each call at a particular instant of time is taken as state '*s*' and is associated with two metrics namely $g = \{delay, loss\}_{avg}$ and $h =$

*{delay$_{est}$, loss$_{est}$}$_{avg}$*. '*g*' calculates the average of delay and loss for already generated states, as measured by network monitoring tool. '*h*' estimates delay (*delay$_{est}$*) and loss (*loss$_{est}$*) for the state to be generated following implementation of QoS mechanism. The state-space scenario for each call is shown in Figure 1.

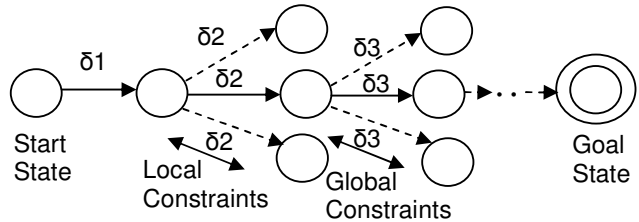

**Index**

| State-Space | Significance In VoIP |
| --- | --- |
| Start State | Call Initiation |
| Goal State | Call Termination |
| Local Constraints | Constraints for each call |
| Global Constraints | Constraints in multiple call scenario |
| δ1 | Change in Network Or Heuristics |
| δ2 | Optimization Technique in single call |
| δ3 | Optimization Technique in multiple call scenario |
| ⟶ | Best ranked action |
| ---▶ | Other actions |

Figure 1. State Space Diagram for the proposed approach

### 3.1.1. Analysis Phase

This phase analyzes all possible conditions of network with respect to delay and loss. There can be 4 scenarios that include delay and loss within tolerable limits, worsening of either delay or loss and finally degradation of both. For each scenario, order of implementation of available QoS mechanisms is selected based on its expected performance. Each such mechanism is denoted by action '*a*'. Mathematically, it is represented by (1) [21].

$$f = \arg\min_{a \in A(s)} h(succ(s,a)) \qquad (1)$$

Successor *succ* is the next state generated due to application of action '*a*' on state '*s*'. *A(s)* denotes set of available actions to optimize state '*s*'. The best ranked action '*a*' is such that by implementing it, '*h*' becomes minimum as denoted by *argmin* function. There are two other functions namely, *one-of (f)* that describes selection of an action from set of suitable actions and *next (f)* that describes selection of next ranked action from set of suitable actions.

### 3.1.2. Implementation Phase

As the call starts, initial state is generated with heuristics. With variation in network parameters or significant change in heuristics, new intermediate states are created. The transition function is termed as *δ1*. Each state is monitored to check whether local constraints are satisfied. Each constraint has a '*threshold*'. If the local constraints are violated, *δ2* is applied to bring the heuristics within threshold. This implies that the best ranked mechanism is applied as per analyzed results. New states are monitored. If local constraints are still not satisfied, next ranked action is implemented and so on.

In multiple call scenarios, global constraints must also be satisfied. Calls with low QoS metrics are classified as '*degraded*' calls and the rest as '*accepted*' calls. Existing QoS implementations for accepted calls are stopped temporarily and are redirected to the degraded calls. As global constraints are satisfied, new states are generated and corresponding transition functions are termed as *δ3*. All these states are monitored again and QoS mechanisms are implemented to satisfy local constraints. High priority calls are given weights and global constraints are calculated as weighted mean of local constraints. This approach is represented in Figure 2.

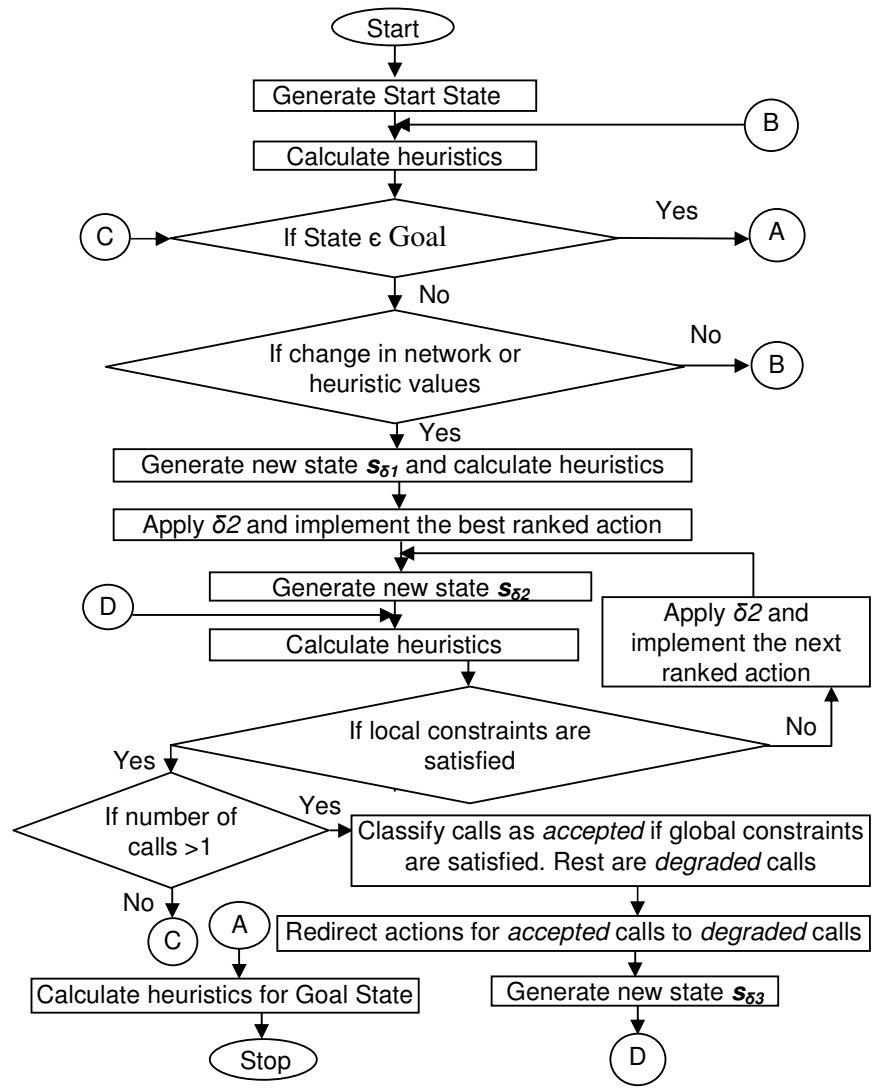

Figure 2. Flowchart depicting the proposed approach

The pseudo-code is given under.

1. $s:=s_{start}$.                          //Call initiation state with heuristics.
2. Calculate $g_s$.                    //Delay, loss measured for current state
3. IF $s \in G$ THEN GOTO step 18.    //Goal state is reached.
4. IF $g_s > threshold$ THEN GOTO step 5 ELSE GOTO step 2.
5. $s:=s_{\delta 1}$. Calculate $g_s$.          //New state is generated.

215

6. *a*:=*one-of(argmin$_{a \epsilon A(s)}$h(succ(s,a))*.  //Select best ranked action.
7. Execute action *a*.                //Action 'a' is implemented.
8. *s*:=$s_{\delta 2}$.                //New state is generated after execution of action 'a'.
9. Calculate $g_s$.                   //Delay, loss measured for current state
10. IF $g_s$<*threshold* THEN GOTO step 13 ELSE GOTO step 11. /*Local constraints must be satisfied.*/
11. *a*:=*next(argmin$_{a \epsilon A(s)}$h(succ(s,a))*.   //Select next action.
12. Execute action *a*. GOTO step 8.
13. IF no. of calls>1 THEN GOTO step 14 ELSE GOTO step 3.
14. Classify ongoing calls as *accepted* calls whose $g_s$<*threshold*. Rest of the calls is *degraded* calls.
15. Stop action *a* ϵ *A(s)* for *accepted* calls.
16. Execute action *a*:=*argmin$_{a \epsilon A(s)}$h(succ(s,a))* in degraded calls.
17. *s*:=$s_{\delta 3}$. GOTO step 9.        //New state is generated after action 'a'
18. Calculate $g_s$ for *s* ϵ *G*.        //Goal heuristics calculated.

### 3.2. Implementation of the algorithm

#### 3.2.1. Description of the test-bed

Our experimental test-bed, as shown in Figure 3 consists of fixed and mobile nodes for VoIP communication, wireless access points, a switch and a Session Initiation Protocol (SIP) server. X-Lite [26] is used as the softphone with support for various audio codecs and Group QoS (GQoS). The Brekeke SIP server [27] is SIP Proxy Server and Registrar. ManageEngine VQManager [28] is used to analyze QoS metrics of ongoing call. Further, User Datagram Protocol (UDP) is used with Real-time Transport Protocol (RTP) on top of it.

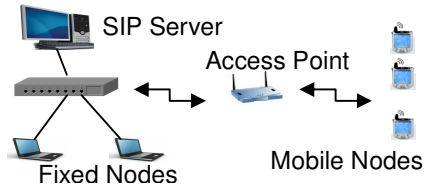

Figure 3. Experimental Test-Bed

#### 3.2.2. Test-bed Analysis Phase

Initially the effect of buffer size is studied. Four scenarios are created using Network Emulator for Windows Toolkit (NEWT) [29] as shown in Table 1. As seen from Table 2, increase in loss rate in network results in degraded performance in terms of packet loss in scenario 4. As buffer size is increased, end-to-end delay increases and retransmissions take place after certain timeout, resulting in further loss as in scenario 3. Even in absence of loss rate, increasing buffer size increases end-to-end delay while decreasing it increases loss as seen in scenarios 1 and 2 respectively. So selection of proper buffer size is important.

Table 1. Different Network Scenarios

| Parameters | Scenario 1 | Scenario 2 | Scenario 3 | Scenario 4 |
|---|---|---|---|---|
| Network Latency in ms | 100 | 100 | 100 | 100 |
| Network Loss in % | Nil | Nil | 30 | 30 |
| Buffer Size in packets | Maximum | 20 | Maximum | 20 |

Table 2. Delay and Loss in each scenario

| Scenarios | Max. Delay | Avg. Delay | Max. Loss | Avg. Loss |
|---|---|---|---|---|
| 1 | 156 ms | 112 ms | 0 % | 0 % |
| 2 | 39 ms | 11 ms | 44 % | 5 % |
| 3 | 94 ms | 6 ms | 61 % | 22 % |
| 4 | 6 ms | 1 ms | 49 % | 21 % |

It is also observed that if the in/out bit rate in an endpoint (caller/callee) varies significantly with time, call quality drops and terminates at last. BroadVoice 32 (BV32) [30] is used as the codec and the in and out buffer size of an end-point is varied to get the results as shown in Table 3. Figure 4(b) shows the call termination as the in/out bit rate varies in contrast to Figure 4(a) where the call continues. Thus it can be inferred that similar buffer sizes and hence comparable bitrates must be maintained for a call to continue successfully.

Table 3. Readings for the variable in/out bit rate in an endpoint

| Local-> Remote Buffer size (packets) | Remote-> Local Buffer size (packets) | Sending Rate (kbps) | Receiving Rate (kbps) | Call Duration (sec) |
|---|---|---|---|---|
| 20 | 20 | 26 | 26 | 1137 |
| 90 | 20 | 26 | 54 | 168 |

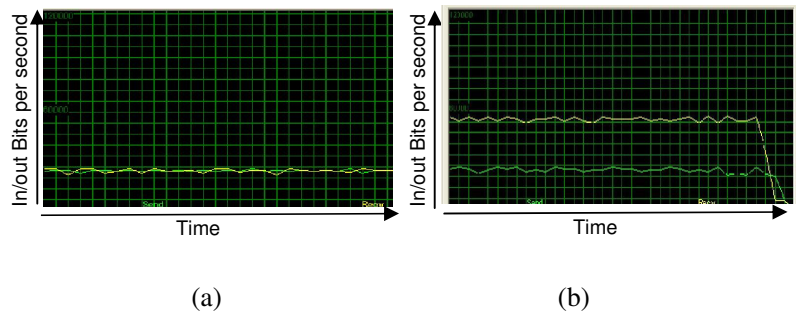

(a)                                       (b)

Figure 4. Effect of (a) constant in/out bit rate (b) variable in/out bit rate on ongoing call

Further, the effect of active queue management is studied. Active queues drop packets before queue is full based on certain probabilities and threshold parameters to maintain bursts in flows and fairness among users. Here Random Early Detection (RED) [31] queue is implemented by keeping maximum threshold at 100 and minimum threshold at 50. We create two congested media having 1kbps and 10 kbps Constant Bit Rate (CBR) background traffic. As observed from Table 4, in moderately congested medium, delay and loss are within tolerable limits. Thus it is advantageous than having fixed size buffer. However, increase in congestion increases loss when RED is implemented. So selection of active queue management policy is of utmost importance towards maintaining quality of call.

Table 4. Different Execution Scenarios of RED implementation

| Parameters | Background traffic 1 kbps | Background traffic 10 kbps |
|---|---|---|
| Average Delay in ms | 66 | 70 |
| Average Loss in % | 6 | 14 |

Finally we implement IntServ model to optimize VoIP performance. IntServ model proposes 2 service classes [32] namely,
1. Controlled load service [33] for reliable and enhanced best-effort service,
2. Guaranteed load service [34] for applications requiring a fixed delay bound.

Both are implemented in the test-bed for the scenarios as mentioned in Table 1. Experiment results conclude that controlled load service gives better performance in terms of packet loss than guaranteed service in scenarios 1, 2 and 3 as seen in Figure 5. In scenario 4 which is the most congested scenario, call terminates in 58 seconds and 111 seconds under guaranteed service and controlled load service respectively. So it is concluded that controlled load service is more suited during congestion than guaranteed service.

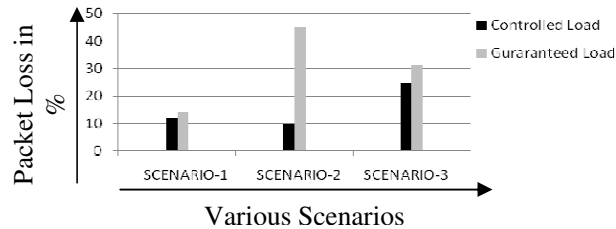

Figure 5. Effect of Controlled Load and Guaranteed Load on loss in various scenarios

### 3.2.2.1. Selection of the order of implementation techniques

From analyzed results, the order of implementation is proposed.
- **Case 1**: Both delay and loss are within tolerable range.
  Guaranteed load service is applied to further enhance it.
- **Case 2**: Delay is tolerable but loss is high.
  Buffer size of access points is increased till acceptable value of delay. Further, RED is applied as the next option. If loss persists, third option is to apply FEC technique. Lastly, controlled load service is applied.
- **Case 3**: Loss is less but delay is high.
  The buffer size of access points is decreased till acceptable value of loss. Weighted RED is applied as the next option with random drop type to ensure fairness. Controlled load service is applied as the last option.
- **Case 4**: Both delay and loss are poor.
  Controlled load service is applied. RED is implemented with small difference between maximum and minimum thresholds as next option.

### 3.2.3. Test-bed Implementation Phase

Our proposed approach is initially implemented in a single call scenario in the test-bed as described in Section 3.2.1. Network conditions are varied using NEWT. Heuristic categories are described in Table 5. Figure 6 shows the state-space diagram for the call. Heuristics for each state are shown in Table 6 and the transition function for every link is described in Table 7.

Table 5. Category of Heuristics

| Heuristic Category | Description |
|---|---|
| Excellent | Delay<=100 ms, Loss <=1%, MOS>=4 |
| Good | 100ms<Delay<=150 ms,1%<Loss<=2%, 3.5<=MOS<4 |
| Average | 150ms<Delay<=180 ms,2%<Loss<=5%, 2<=MOS<3.5 |
| Poor | Delay>180 ms, Loss>5%, MOS< 2 |

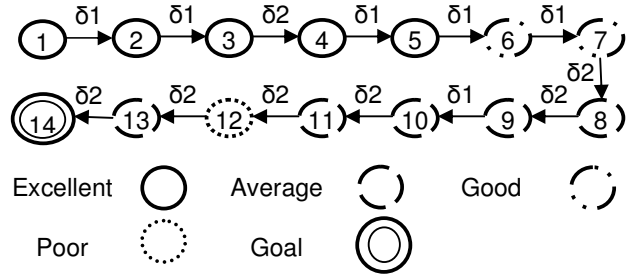

Figure 6. State transition diagram for the call

Table 6. Heuristics for each state during the call

| State | Delay (ms) | Loss(%) | MOS | Duration (s) | Comments |
|---|---|---|---|---|---|
| 1 | 6 | 0 | 4.4 | 1 | Start State |
| 2 | 19 | 0 | 4.4 | 420 | Excellent |
| 3 | 85 | 0 | 4.4 | 530 | Excellent |
| 4 | 69 | 0 | 4.4 | 90 | Excellent |
| 5 | 95 | 0 | 4.4 | 350 | Excellent |
| 6 | 131 | 0 | 4.4 | 137 | Good |
| 7 | 147 | 0 | 4.4 | 126 | Good |
| 8 | 169 | 0 | 4.4 | 600 | Average |
| 9 | 164 | 0 | 4.4 | 187 | Average |
| 10 | 160 | 2 | 3.3 | 143 | Average |
| 11 | 169 | 2 | 2 | 136 | Average |
| 12 | 169 | 2 | 1.8 | 136 | Poor |
| 13 | 143 | 2 | 2 | 340 | Average |
| 14 | 163 | 2 | 2 | 250 | Goal State |

Table 7. Transition function for every link between the states

| Link | Transition Functions | Comments |
|---|---|---|
| 1-2 | 50 ms latency($\delta 1$) | |
| 2-3 | 65 ms latency($\delta 1$) | |
| 3-4 | Guaranteed service applied($\delta 2$) | Delay decreases to some extent |
| 4-5 | 80 ms latency($\delta 1$) | |
| 5-6 | 120 ms latency($\delta 1$) | |
| 6-7 | ($\delta 1$) | Delay varies significantly |
| 7-8 | Buffer size reduced to 50($\delta 2$) | Delay decreases |
| 8-9 | Buffer size reduced to 30($\delta 2$) | Delay reaches threshold mark. |
| 9-10 | 0.01 Loss rate($\delta 2$) | 1 out of every 100 packets is lost. |
| 10-11 | Buffer size increased to 45($\delta 2$) | To decrease loss & enhance MOS |
| 11-12 | Buffer size increased to 60($\delta 2$) | It is done to decrease increasing loss. MOS now becomes uniform. |
| 12-13 | RED applied with max threshold of 100 and min threshold of 50($\delta 2$) | As buffer size cannot be increased further due to increase in delay, the next best ranked action is chosen. |
| 13-14 | Controlled load is applied($\delta 2$) | MOS gets improved. |

The average delay is 120 ms and packet loss is 1%. The average MOS is 3.3. Thus the call is of acceptable quality. Readings from VQManager as seen in Figure 7(a) suggest that loss and delay remain uniform and tolerable in degraded conditions.

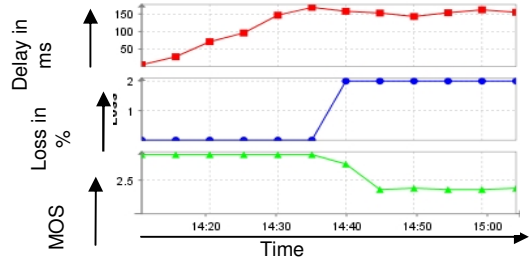

(a)

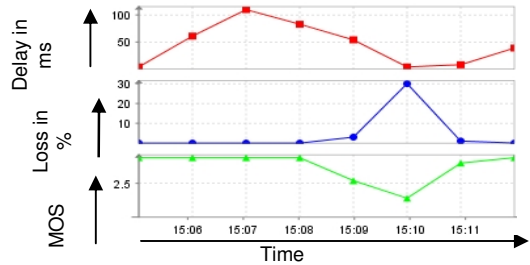

(b)

Figure 7. Variation of delay, loss and MOS in (a) single and (b) multiple call scenario

The proposed algorithm is now implemented in a multiple call scenario. 2 calls are made simultaneously between two soft phones and between a mobile phone and a soft phone respectively. These calls are mapped as their state-space diagrams. It is seen from Figure 7(b) that as overall delay of the calls increases, the buffer size is decreased as per $\delta 2$ satisfying the local constraints. However, overall loss for all the calls increases. So $\delta 3$ is implemented following the proposed algorithm by increasing the buffer size to satisfy the global constraints. While this increases the delay again, it is tolerable. On the other hand, loss decreases and MOS remains at a tolerable limit. The maximum, minimum and average heuristic values are shown in Table 6.8.

Table 8. Heuristic values in multiple call scenario

| Parameters | Minimum | Maximum | Average |
|---|---|---|---|
| Delay (ms) | 4 | 110 | 49 |
| Loss (%) | 0 | 30 | 5 |
| MOS | 1.4 | 4.4 | 3.6 |

Finally the algorithm is applied to maintain the QoS of a video call. An average video call in Internet incurs a loss of 1-2%. Network loss is therefore introduced to degrade the video call in the test-bed. Figure 8(a) and Figure 8(b) show the quality of the received picture under 0% and 3 % packet loss respectively. As loss reaches 4 %, the degradation is quite noticeable as shown in Figure 8(c). The picture quality slowly becomes unrecognizable as loss reaches 6% as seen from Figure 8(d). After implementation of the proposed algorithm, loss is again tolerable as it reaches the threshold limit of 5 % and the video becomes recognizable in the degraded scenario as can be inferred from Figure 8(e). Thus the proposed algorithm helps to maintain the QoS of the video call adaptively even as loss increases in the network.

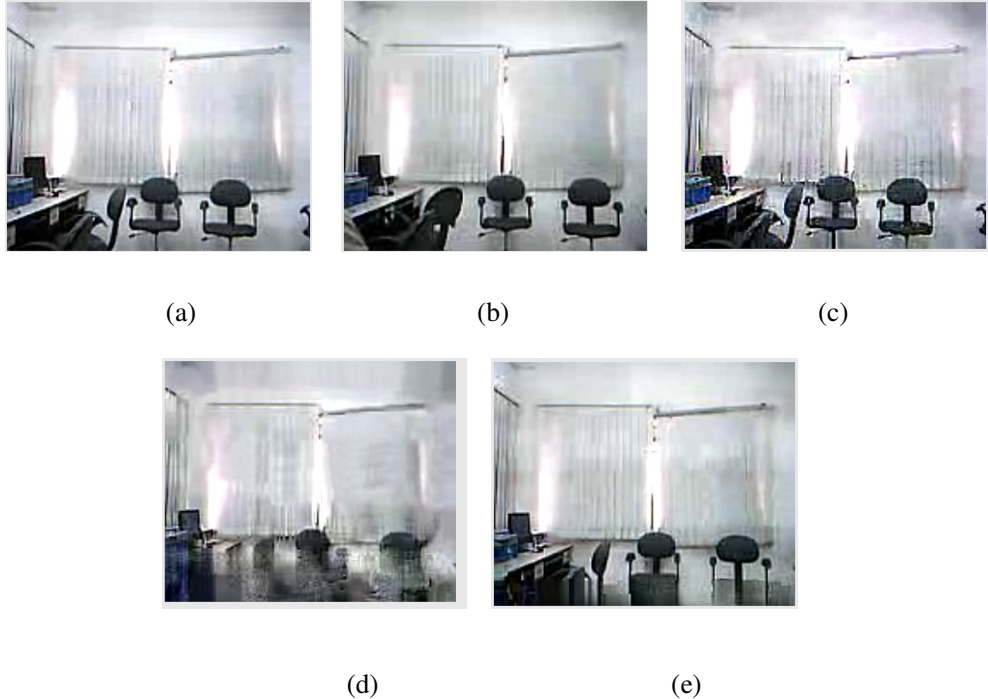

Figure 8. Images during a video call with (a) 0 % (b) 3 % (c) 4 % (d) 6 % (e) 5 % loss

## 4. IMPLEMENTATION OF LEARNING STRATEGY FOR QOS ENHANCEMENT

It is observed from Section 3 that the QoS implementation techniques are repeated in the same order every time for each scenario. Moreover as learnt from the implementation results in Section 3.2.3, the best ranked mechanism may not be always suitable in a particular scenario. Hence the next ranked action is selected from the knowledge base and this process continues until the most suitable action is found that maintains both delay and loss within tolerable limits. Thereafter, the knowledge base is not updated and every time the same sequence of actions has to be repeated for reaching the acceptable call quality with respect to a particular scenario. Our objective is, therefore to update the knowledge base so that the most suitable QoS implementation mechanism is applied as soon as possible. Therefore, the idea of "learning" from the environment is applied to the existing algorithm.

Learning is the acquisition of new declarative knowledge, the development of motor and cognitive skills through instruction or practice, the organization of new knowledge into general, effective representations, and the discovery of new facts and theories through observation and experimentation [35]. Our aim is to make the algorithm "learn" from the environment as the call goes on and accordingly update the knowledge base. However, inclusion of learning must also satisfy our initial objective as mentioned in Section 2. The primary advantage of incorporating learning to the existing incremental heuristic search is to make the algorithm more dynamic and adaptive to the changing scenarios.

## 4.1. Proposed Algorithm

The "learning" algorithm is implemented in conjunction with the previous algorithm as stated in Section 3. This algorithm comprises of two sections namely knowledge acquisition and skill refinement. They are described in the following subsections.

### 4.1.1. Knowledge Acquisition Phase

Knowledge acquisition is defined as learning new symbolic information coupled with the ability to use that information in an effective manner [35]. The essence of knowledge acquisition in this algorithm is building the knowledge base to explain broader scope of situations, to predict the behaviour of the physical world and finally to be more accurate in application of suitable actions. The knowledge base is initially built during the analysis phase as described in Section 3.1.1. The full knowledge base is finally developed with the steps described below.

1. Associate each action '$a$' with its '$h$' value.
2. Rank '$a$' based on the category of heuristics in which '$h$' lies. *one-of (f)* denotes that all '$a$' have the same rank. *next (f)* denotes that '$a$' has the next lower rank.
3. Calculate $g(s)$ as '$a$' is applied to any particular state '$s$'.
4. Update $h_a = g(s)$.

### 4.1.2. Skill Refinement Phase

Skill refinement is defined as the gradual improvement of skills by repeated practice and correction of deviations from the desired behaviour [35]. Skill refinement is used in this algorithm in order to update the knowledge base in the run-time as the call goes on. Since the analysis phase as described in Section 3.1.1 is performed in a particular environment based on predictions and certain estimations, the knowledge base built thereafter may have certain discrepancies that are detected only in the run-time. Skill refinement phase of the 'learning' algorithm helps to eradicate these erroneous attributes of the knowledge base so that the most suitable action can be applied in the least possible time. While knowledge acquisition is the initial stage in developing the knowledge base, skill refinement focuses on updating the knowledge and is described in the following steps.

Let '$a_{current}$' be the final action that satisfies the local constraints after implementation on the ongoing call at the current point of time. Let '$C$' contains the set of conflicting actions denoted by $C = \{S_1, S_2, \ldots S_n\}$
where $S_1 = (a_1, a_2, \ldots, a_n)$ = one set of conflicting actions,
$S_n = (a_{11}, a_{22}, \ldots a_{nn})$ = another set of conflicting actions, and so on.

```
1.  FOR all actions a of a particular scenario such that rank(a) > rank(a_current)
2.  {
3.          IF h(a) > h(a_current)
4.          {
5.                  IF (( a and a_current) ∈ S_i for i = 1 to n)
6.                  {
7.                          rank(a_current) = rank(a)
8.                          delete a
9.                  }
10.         ELSE
11.                 {
12.                         swap (rank(a_current) , rank(a))
13.                 }
14.         }
15. }
```

## 4.2. Implementation of the Algorithm

The knowledge base for appropriate ranking and implementation of the best ranked action is initially developed during the analysis phase as described in Section 3.2.2.1. The knowledge base is further updated by the "learning" algorithm as described in Section 4.1. The algorithms as described in Section 3.1 and Section 4.1 respectively are implemented in a single call under varying network scenarios as set in Section 3.2.3. The test-bed as described in Section 3.2.1 is used for implementation. It is seen from Figure 9 that delay decreases and MOS gets enhanced with respect to time after implementation of the algorithms.

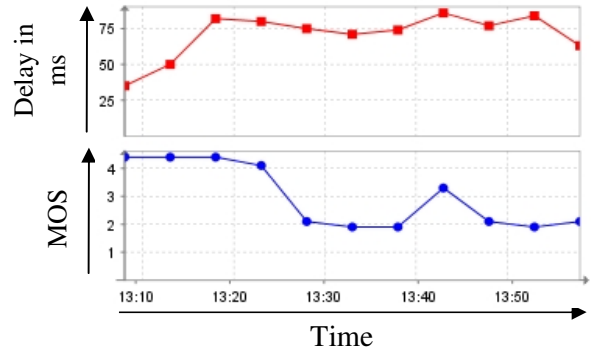

Figure 9. Variation of delay and MOS with time

The effect of the learning algorithm is clearly visible with respect to packet loss in Figure 10. In region 1, loss rate is increased in the network. Actions are implemented as per their ranking in the knowledge base till loss is minimized below the threshold level. This entire process has consumed approximately 10 minutes. In effect, the initial algorithm as described in Section 3.1 is implemented. The network parameters are reset and again configured to same values in region 2. With the learning algorithm in execution, the knowledge base is now refined. The result is quite evident in region 3 where the most suitable action is applied in the least possible time, thereby preventing any increase in loss unlike the previous scenario. In other words, the algorithm learns from its past experience and updates itself accordingly. As seen from Figure 7(a) and Figure 9, while the average MOS values are comparable, MOS improves to 2.1 in Section 4.2 compared to MOS value of 2 in Section 3.2.3 clearly reflecting the fact that introduction of learning has further enhanced the call quality under diverse and degraded scenarios.

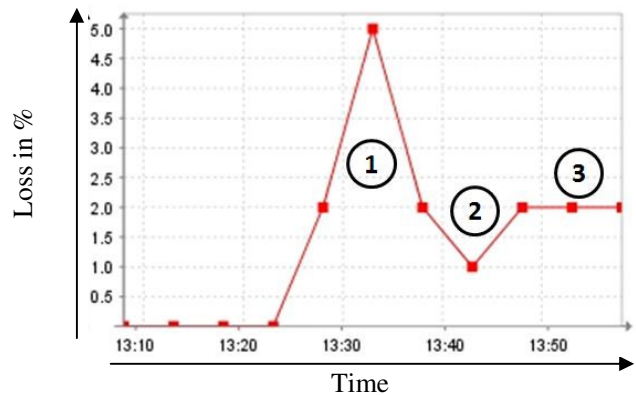

Figure 10. Variation of packet loss with time

## 5. BENEFITS

The benefits of the proposed algorithms are discussed with respect to state-space search and real-time traffic. Our proposed approach addresses the issue of QoS maintenance in real-time applications like VoIP by mapping it into a state-space search problem. This mapping is advantageous as it can be further optimized by applying advanced search techniques. In addition, new optimizations, for instance as suggested in [36], can also be applied. Current VoIP developers like CISCO implement QoS using dedicated systems for node management, link management and traffic management. Such systems may conflict with each other unless they are applied in proper sequence. Our approach of application of QoS implementation techniques in proper sequence guarantees an improved and simplified performance.

Implementation of the state-space search makes the algorithm adaptive to changing network scenarios and satisfies our objective as discussed in Section 2. Introduction of learning strategy in the algorithm makes the algorithm dynamic and more intelligent. While performance improvement supports this claim, advanced forms of learning, for instance, as described in [37] and [38] can be further used for increasing the efficiency of action implementation.

As QoS is the main area of focus, its underlying principles must be followed. Firstly, applications must be shielded from the complexities of the underlying QoS specification and management [39]. Our approach aims to build an automated system which will apply the transition functions to satisfy constraints. Thus it helps in relieving the VoIP applications from the complexities of QoS maintenance, thereby maintaining transparency. Moreover, our algorithm helps to make QoS configurable, predictable and maintainable [39].

Viewing from the state-space search perspective, our proposed solution satisfies the following criteria.

- Inferential adequacy - Since any QoS mechanism can be fed into the transition function after proper analysis and ranking, our approach will never fall short of providing QoS under varying network scenarios.

- Inferential efficiency - Our proposed algorithm adaptively maintains the calls within acceptable delay and loss limits. While each call remains satisfactory, the global status of all the calls is also tolerable. Thus efficiency is achieved.

- Aquisitional efficiency - Existing optimization techniques can be incorporated into the proposed algorithm. For example in our implementation, as wireless calls are being made, parameter optimizations of APs as described in [40] can also be implemented. Moreover, techniques for enhancing VoIP performance under congested scenarios can also be incorporated in this work, for example, by varying the packet payload size as suggested in [41]. Advanced artificial intelligence search techniques can be further applied.

## 6. CONCLUSION

In this paper, we have dealt with the problem of adaptive QoS maintenance under dynamic and diverse network conditions and applied optimization techniques accordingly. Test-bed readings verify the fact that application of the proposed algorithms in single and multiple voice and video calls keeps both delay and loss within threshold limits even as network conditions vary with respect to time. The algorithms further ensure that no conflict arises during the application of optimization techniques as proper sequence is maintained among them. Performance improvement is observed after introduction of learning as it refines the knowledge base and makes the algorithms more efficient in terms of application of the most suitable QoS implementation technique in the least possible time under diverse network scenarios. While VoIP traffic binds these algorithms to real-time heuristic search, modern optimizations in this dynamic search domain can be further applied to the state space search approach.

## ACKNOWLEDGEMENTS

The authors deeply acknowledge the support from DST, Govt. of India for this work in the form of FIST 2007 Project on "Broadband Wireless Communications" and in the form of PURSE 2009 Project on "Cognitive Radio" in the Department of ETCE, Jadavpur University.

International Journal of Wireless & Mobile Networks (IJWMN) Vol. 3, No. 5, October 2011

**Authors**

**Tamal Chakraborty** (tamalchakraborty29@gmail.com) has done Master of Technology in Distributed and Mobile Computing from Jadavpur University, Kolkata, India. He has studied Computer Science and Engineering from Future Institute of Engineering and Management under West Bengal University of Technology, Kolkata, India and received his Bachelor of Technology Degree in 2009. After the completion of his M.Tech, he has joined the Department of Electronics & Tele-communication Engineering, Jadavpur University as a Senior Research Fellow. His research interests include Voice over IP and Wireless Communication.

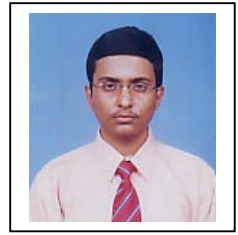

**Atri Mukhopadhyay** (atri.mukherji11@gmail.com) has done Master of Technology in Distributed and Mobile Computing from Jadavpur University, Kolkata, India. He has studied Electronics and Communication Engineering from Meghnad Saha Institute of Technology under West Bengal University of Technology, Kolkata, India and received his Bachelor of Technology Degree in 2009. His research interests focus on Voice over IP and Wireless Communication.

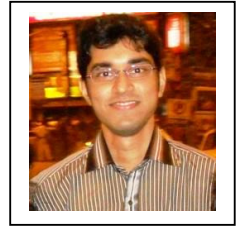

**Dr. Iti Saha Misra** (itisahamisra@yahoo.co.in; iti@etce.jdvu.ac.in) is presently holding the post of Professor in the Department of Electronics and Telecommunication Engineering, Jadavpur University, Kolkata, India. After the completion of PhD in Engineering in the field of Microstrip Antennas from Jadavpur University (1997), she is actively engaged in teaching since 1997.

  Her current research interests are in the areas of Mobility Management, Network

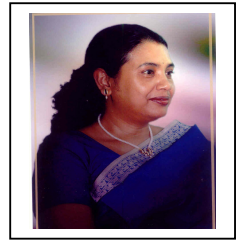

Architecture and protocols, Integration Architecture of WLAN and 3G Networks, Call Admission control and packet scheduling in cellular and WiMAX networks. Her other research activities are related to Design Optimization of Wire Antennas using Numerical Techniques like GA, PSO and BFA. She has authored more than 100 research papers in refereed Journals and International Conferences and published a Book on Wireless Communications and Networks, by McGraw Hill.

She is the recipient of the Career award for Young teachers by All India Council for Technical Education (AICTE) for the financial year 2003-2004 and obtained the IETE Gowri memorial award in 2007 for being the best paper in the general topic of 4G networks: Migration to the Future. She has developed the OPNET, QualNet and VoIP laboratories in the Department of Electronics and Telecommunication Engineering of Jadavpur University to carry out advanced research work in Broadband wireless domain.

Mrs. Saha Misra is the Senior Member of IEEE and founder Chair of the Women In Engineering, Affinity Group, IEEE Calcutta Section.

**Dr. Salil K. Sanyal** (s_sanyal@ieee.org) received his Ph.D (Engineering) Degree from Jadavpur University, Kolkata – 700032, West Bengal, India in 1990. He joined the Department of Electronics and Telecommunication Engineering, Jadavpur University as Lecturer in 1982 where currently he holds the post of Professor. He is the immediate past Head of the Department of Electronics and Telecommunication Engineering of Jadavpur University.

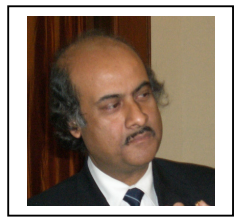

He has published more than 130 Research Papers in reputed peer reviewed Journals and International/National Conference Proceedings.

Mr. Sanyal is a Senior Member of IEEE and also the past Chair of IEEE Calcutta Section. His current research interests includes Analog/Digital Signal Processing, VLSI Circuit Design, Wireless Communication and Tunable Microstrip Antenna.